\documentclass[12pt,a4paper,twoside,groupcitations]{article}
\usepackage[T1]{fontenc}
\usepackage[ansinew]{inputenc}
\usepackage[english]{babel}
\usepackage{amsfonts}
\usepackage{amsmath}
\usepackage{array}
\usepackage{amsthm}
\usepackage{amssymb}
\usepackage{graphicx}
\usepackage{subfigure}
\usepackage{braket}
\usepackage{cite}
\usepackage{verbatim}
\usepackage[table]{xcolor}
\usepackage{caption}
\usepackage{cite}
\usepackage{textcomp}
\usepackage{bm}
\usepackage{mathtools}
\usepackage{multicol}
\usepackage{tikz}
\usetikzlibrary{positioning,arrows}
\usetikzlibrary{decorations.pathmorphing}
\usetikzlibrary{decorations.markings}
\usetikzlibrary{calc,decorations.markings}
\usetikzlibrary{arrows,shapes}
\usetikzlibrary{matrix,arrows}
\usepackage{xparse}
\definecolor{jade}{HTML}{00A86B}
\newcommand{\be}{\begin{eqnarray}}
\newcommand{\ee}{\end{eqnarray}}

\renewcommand{\d}{\mbox{${\rm d}$}} 

\newcommand{\beq}{\begin{equation}}
\newcommand{\eeq}{\end{equation}}

\DeclareMathOperator{\Det}{Det}
\title{\bf Background independence and field redefinitions in quantum gravity}
\author{Roberto~Casadio$^{ab}$\thanks{E-mail: casadio@bo.infn.it},
$\ $
Alexander~Kamenshchik$^{ab}$\thanks{E-mail: kamenshchik@bo.infn.it},
$\ $
and
Iber\^e Kuntz$^{c}$\thanks{E-mail: kuntz@fisica.ufpr.br}
\\
\\
$^a${\em Dipartimento di Fisica e Astronomia, Universit\`a di Bologna}
\\
{\em via Irnerio~46, 40126 Bologna, Italy}
\\
\\
$^b${\em I.N.F.N., Sezione di Bologna, I.S.~FLAG}
\\
{\em viale B.~Pichat~6/2, 40127 Bologna, Italy}
\\
\\
$^c${\em Departamento de F\'isica, Universidade Federal do Paran\'a}
\\
{\em PO Box 19044, Curitiba -- PR, 81531-980, Brazil}
}
\begin{document}
\date{}
\maketitle
\begin{abstract}
It is generally believed that a full-fledged theory of quantum gravity should exhibit background independence and diffeomorphism
invariance.
In its most general form, the latter comprises field redefinitions, which are diffeomorphisms in configuration space.
We show that any path-integral approach to quantum gravity leads to a tension between these properties, such that they cannot hold
simultaneously.
\end{abstract}
%
%
%
%
%
%
%
\section{Introduction}
\label{sec:intro}
\setcounter{equation}{0}
Background independence and covariance are the basic building blocks of general relativity.
Both concepts are intertwined with the diffeomorphism invariance of gravity and has led to a long history
of confusion~\cite{Norton_1993,Pooley:2015nfa,Anderson}.
Anderson~\cite{Anderson} was the first to put such concepts into a more precise language by demarcating
the differences of absolute and dynamical structures.
Although not free of inconsistencies~\cite{Giulini:2006yg}, Anderson's approach delineates the role played by
the diffeomorphism group in gravitational and non-gravitational theories, which is a stepstone for defining
covariance and background-independence rigorously.

These concepts are also believed to play a fundamental role in the quantization of gravity~\cite{Colosi:2004sa,Smolin:2005mq}.
The conventional perturbative approach of quantum field theory singles out a background metric required
for the vacuum structure of the theory.
Because this background is not dynamical, it violates the background independence of general relativity.
Allowing for this possibility, but insisting on the invariance under diffeomorphisms, is the cornerstone of
string theory~\cite{Polchinski:1998rq,Polchinski:1998rr,Witten:1993ed,Rozali:2008ex}.
On the other hand, Canonical Quantum Gravity and some of its variants,
including Loop Quantum Gravity,
has been developed as a background-independent theory
\textit{ab initio\/}~\cite{DeWitt:1967yk,Bergmann:1966zza,Dirac:1958sq,kiefer,Rovelli:2004tv,Ashtekar:2004eh}.
Nonetheless, the absence of spacetime structures makes it difficult to recover well-tested results of quantum field
theory at low energies.
\par
Quantum field theories of gravity can also be formulated, to some extent, in a background-independent manner.
This is indeed implemented with the background field method~\cite{Abbott:1981ke}, where a separation of quantum
perturbations from (dynamical) background fields is performed.
The background field method was introduced as a convenient way to deal with gauge theories, in such a way that
the gauge symmetry could be preserved at the background level while fixing the gauge for the quantum perturbations.
The resulting action is therefore gauge invariant, which eases tremendously some calculations, such as the proof
of renormalizability.
\par
The background field, despite its name, need not be chosen \textit{a priori\/}.
It need not even satisfy the classical equations of motion.
Indeed, the background field should solve the effective equations of motion.
It is in this sense that a quantum field theory of gravity can be background independent~\cite{reuter}.
Nonetheless, this construction requires the definition of the asymptotic vacuum state, thus restricting the set
of background metric solutions to the ones with sufficient asymptotic symmetry.
We stress, however, that it is the asymptotic boundary conditions rather than the background metric that are
fixed beforehand.
\par
When the background field is not chosen as a solution to the field equations, the effective action turns out
to depend on the gauge.
One thus reaches an impasse, where either a background is picked up \textit{a priori\/}, violating the
background independence of general relativity but whose predictions are gauge-independent, or the background
is solved for \textit{a posteriori\/}, which respects background independence but suffers from gauge dependence.
This issue is solved within the Vilkovisky-DeWitt formalism \cite{Vilkovisky:1984st,DeWitt:1988dq},
where a Levi-Civita connection in configuration space is introduced to cancel out the gauge-dependence while
maintaining the background fields (including the metric) arbitrary.
The cancellation of the gauge dependence also extends the invariance of on-shell scattering amplitudes under field
redefinitions to off-shell correlation functions.
We note that field redefinitions are nothing else than diffeomorphisms in configuration space.~\footnote{An example of field redefinitions often treated in gravity and cosmology is the transition between the Jordan frame and the Einstein frame. It is important to emphasize that this transition includes not only the conformal transformation of the metric turning the nonminimal coupling between the scalar field and the curvature into the minimal one, but also the redefinition of the scalar field to make its kinetic term canonical \cite{Wagoner}. The question of the classical and quantum equivalence between two frames was extensively discussed in the literature (see e.g.~\cite{Kam-Stein} and \cite{Falls:2018olk} and the references therein). We think that these frames are equivalent mathematically, in the sense that one can describe everything using any frame (just like one can use inertial and non-inertial frames in Newtonian mechanics). At the the same time, physically, observers connected with different frames can see different phenomena. For example, in one frame one can have de Sitter expansion, while in another frame one sees other law of the expansion or even some kind of bounce~\cite{Gianni}.} In the remainder of this article,
covariance shall thus mean both spacetime and configuration-space covariance.
\par
The purpose of this paper is to show that quantum field theories of gravity, even within the formalism of Vilkovisky-DeWitt,
cannot consistenly satisfy both background-independence (absence of \textit{any} non-dynamical structure) and covariance
(under spacetime diffeomorphisms and field redefinitions) at the same time.
As we shall show, the latter necessarily requires absolute structures to be introduced, which in turn violates the former,
whereas a theory that satisfies the former does not possess the correct transformation properties, thus violating the latter.
\section{The geometrical effective action}
\label{sec:EA}
\setcounter{equation}{0}
A valuable property of quantum field theory regards the invariance of the S-matrix under field redefinitions~\cite{Rebhan:1987cd}.
Scattering amplitudes are indeed the observables of interest in Particle Physics, thus their invariance under field reparameterizations
seems to suggest some profound principle.
Observables in Astrophysics and Cosmology are, however, correlation functions and these do not transform covariantly
in the standard formulation of quantum field theory.
One could argue that this proves the absence of such a general principle.
Nonetheless, in a gauge theory, field-parameterization dependence induces gauge dependence in the observables.
The former is, therefore, a genuine issue.

The origin of this gauge dependence is most easily seen by the naive effective action
\begin{equation}
	\exp\left(\frac{i}{\hbar}\, \Gamma_\text{N}[\phi]\right)
	=
	\int \mathcal{D}\varphi
	\,
	\exp \left\{\frac{i}{\hbar} \left[
		S[\varphi]
		+ (\varphi^i - \phi^i)\frac{\delta \Gamma_\text{N}}{\delta \phi^i}
		\right]
		\right\}
	\ ,
	\label{EA}
\end{equation}
where $\varphi^i = \varphi^I(x)$ collectively denotes an arbitrary set of fields with classical action $S[\varphi]$ and $\phi^i$
is the mean field.
The set of all $\varphi^i$ is the configuration space, formally defined by the infinite-dimensional space
\begin{equation}
	\mathcal C(M)
	=
	\prod_x \mathcal N(x)
	\ ,
\end{equation}
where $\mathcal{N}(x)$ denotes the finite-dimensional space formed by all $\varphi^I(x)$ evaluated at the same spacetime point $x\in M$. 
Under gauge transformations or field redefinitions, neither the path-integral measure nor the second term within the curly brackets in
Eq.~\eqref{EA} remains invariant.
One can impose invariance in the former by introducing a configuration-space metric $G_{ij}$, whereas the latter calls for a
configuration-space connection $\Gamma^k_{ij}$ compatible with $G_{ij}$.
This procedure results in the Vilkovisky-DeWitt effective action~\cite{Vilkovisky:1984st,DeWitt:1988dq}.
\par
The configuration-space metric $G_{ij}$ is usually identified as the bilinear operator from the classical kinetic
term~\cite{Vilkovisky:1984st}.~\footnote{We recall that the canonical quantisation of metric theories of gravity is based on the Bergmann-Komar group \cite{BK} of the super-momenta and super-Hamiltonian, whose explicit form depends on the spatial part of the DeWitt metric, hence the choice of field dynamics.}
However, this procedure to fix $G_{ij}$ is not required nor physically well-motivated. It is a pure mathematical procedure.
For this reason, we shall see $G_{ij}$ as part of the definition of the theory, together with a choice for the classical action
(but not derived from it).
One can then determine $G_{ij}$ using symmetry and an expansion on the energy scale, in the same spirit of effective
field theories~\cite{Casadio:2021rwj}.
\par
For the sake of simplicity, in this paper we shall omit the connection-dependent term and focus on the
functional measure.~\footnote{For this reason, our argument also applies to any construction of the path integral where
a functional measure is required, like in non-linear sigma models.}
This is justified because of the compatibility property of the aforementioned connection, which then provides no additional
absolute structures (to be defined in Section~\ref{back-cov}) other than the already present metric $G_{ij}$,
thus not altering our argument of Section~\ref{back-red}.
The improved effective action shall then be written as
\begin{equation}
	\exp\left(\frac{i}{\hbar} \Gamma_\text{VD}[\phi]\right)
	=
	\int \mathcal{D}\varphi
	\sqrt{\Det{G_{ij}}}
	\,
	\exp \left( \frac{i}{\hbar} S[\varphi] \right)
	\ ,
	\label{VD-EA}
\end{equation}
where $\Det$ denotes the functional determinant.
\section{Background independence and covariance}
\label{back-cov}
\setcounter{equation}{0}
In this section, we shall review Anderson's efforts~\cite{Anderson} to define covariance rigorously.
In the notations of~\cite{Giulini:2006yg}, a theory's equations of motion are generally of the form
\beq
\mathcal F(\gamma, \phi, \Sigma) = 0
\ ,
\label{eom}
\eeq
where $\gamma$ denotes all structures defined by maps into spacetime (e.g. a particle's trajectory) and $\phi$
denotes collectively all structures represented by maps from spacetime (e.g.~fields).
Here $\Sigma$ is a given non-dynamical structure defined on the spacetime $M$ for which the equations of
motion~\eqref{eom} can be solved for $(\gamma,\phi)$.
A theory is called covariant iff, for every $f\in \text{Diff}(M)$,
\beq
\mathcal F(f\cdot\gamma, f\cdot\phi, f\cdot\Sigma) = 0 \iff \mathcal F(\gamma, \phi, \Sigma) = 0
\ ,
\eeq
where the dot product $f\cdot X$ denotes the action of $f$, by means of pullback/pushforward, on the object $X$.
On the other hand, a theory is said to be invariant iff, for all $f\in \text{Diff}(M)$,
\beq
\mathcal F(f\cdot\gamma, f\cdot\phi, \Sigma) = 0 \iff \mathcal F(\gamma, \phi, \Sigma) = 0
\ .
\eeq
The subtle difference lies on the transformation of the non-dynamical structure $\Sigma$.
Covariance is the requirement for the theory to be well-defined geometrically, i.e.~for the theory to live on $M$.
Thus, when some structure is transformed under a diffeomorphism, all other structures of the theory are transformed
along to counteract the differences and keep their relation the same.
Invariance, on the other hand, requires solutions of the theory to be transformed into other solutions for the \textit{same} $\Sigma$.
The latter is much more restrictive and constitutes an important guiding principle in physics.
\par
The structure $\Sigma$ is called an absolute structure if it is a field that either is not dynamical or whose solutions are all locally
diffeomorphic to each other.
One can then define a background-independent theory as a theory which contains no absolute structure.
Notice that, in this case, covariance equals invariance.
\par
In special relativity, for instance, the absolute structure is the Minkowski metric $\Sigma = \{\eta_{\mu\nu}\}$.
The passage to general relativity is characterized by making the metric dynamical $\eta_{\mu\nu} \to g_{\mu\nu}$.
Now $g_{\mu\nu}$ is not given, thus $\Sigma = \varnothing$, but is found \textit{a posteriori} as a solution of Einstein's field
equations, hence $\phi = \{g_{\mu\nu}\}$.
\par
As we anticipated in the Introduction, Anderson's definitions are not flawless.
To see this, one can write the dynamics of a scalar field in special relativity in a diffeomorphism-invariant form as
\begin{align}
	g^{\mu\nu}\, \nabla_\mu\nabla_\nu \Phi &= 0
	\ ,
	 \\
	R_{\mu\nu\rho\sigma}
	&=
	0
	\ .
	\label{eq:SR}
\end{align}
According to Anderson's definitions, this would also make special relativity background independent,
which is an obvious flaw.
First of all, Anderson's definition of an absolute structure does not entail the action-reaction principle.
While $\eta_{\mu\nu}$ is obtained dynamically, thereby acting on $\Phi$, it is not acted upon by $\Phi$.
A fully dynamical field should have a symmetrical role where it acts on and is also acted upon by other
fields~\footnote{This property is somewhat related to Mach's principle~\cite{Ehlers1995,Einstein1922}.}.
Interactions are indeed always symmetric.
\par
The source of the problem lies in the fact that Eq.~\eqref{eq:SR} is not a dynamical equation,
but rather a constraint. 
In fact, the most general solution $g_{\mu\nu} = \alpha \, \eta_{\mu\nu}$, for any non-zero constant $\alpha$, 
does not contain enough integration constants to fully account for the initial conditions, hence the Cauchy
evolution is ill-defined.
Furthermore, Eq.~\eqref{eq:SR} does not follow from any diffeomorphism-invariant action unless a Lagrange
multiplier is introduced, which stresses the fact that this equation is a constraint.
\par
In view of the above issues, we shall refine the definitions of covariance, invariance and background independence.
Rather than basing our definitions on the equations of motion, our proposal shall be performed at the level of the effective action.
This not only avoids the issues outlined above, but generalises the discussion to the quantum realm, which is our main concern.
Moreover, the equations of motion do not capture important physical effects contained in the boundary terms
(vide black hole entropy).
The effective action, on the other hand, contains information about all physics at classical or quantum scales.
\par
We shall define an absolute structure $\Sigma$ as any local object that does not contain a corresponding kinetic
term in the effective action.
A theory is then called covariant iff, for every $f\in \text{Diff}(\mathcal C(M))$, the effective action transforms as
\begin{equation}
	\Gamma[f \cdot \phi, f \cdot \Sigma]
	=
	\Gamma[\phi, \Sigma]
	\ ,
	\label{eq:cov}
\end{equation}
whereas we say that it is invariant iff, for every $f\in \text{Diff}(\mathcal C(M))$, we have~\footnote{More generally,
a symmetry group $G\subset \text{Diff}(\mathcal C(M))$ of some $\text{Diff}(\mathcal C(M))$-covariant theory is the group
that leaves the absolute structure $\Sigma$ invariant.
An invariant theory is thus invariant under the largest symmetry group $G=\text{Diff}(\mathcal C(M))$.}
\begin{equation}
	\Gamma[f \cdot \phi, \Sigma]
	=
	\Gamma[\phi, \Sigma]
	\ .
	\label{eq:inv}
\end{equation}
Because we are interested in a fundamental theory, where all interactions are entirely described with fields,
we have not included $\gamma$ in our definitions.
We should also stress that, like in Anderson's formulation, $\Sigma$ is really any non-dynamical object,
not necessarily a spacetime object.
A theory is then said to be background independent whenever $\Sigma = \varnothing$.
\par
We stress that some absolute global framework shall always be needed in order to define a workable
mathematical theory of physical phenomena.
For instance, the spacetime $(M, g_{\mu\nu})$ should be globally hyperbolic in order to give rise to well-posed
Cauchy problems in classical general relativity~\cite{Landsman:2022caj}, and similarly in (classical or quantum)
field theories, although such restrictions are sometimes lifted or overlooked.
This means that properties like the dimension, topology, and differential structures of the spacetime (or field space)
must be fixed to some extent from the onset when describing specific systems.
In practical terms, our discussion of background independence will therefore concern the existence of a
fixed (local) metric at some level of mathematical abstraction, as we will further clarify in the following.
Points, dimension, topological and differential structures are therefore not included in our definition
of background independence.
\par
Before we proceed, it is important to recall that observables in general relativity are mathematically given by scalars
built out of (pairs of) tensors pertaining to the quantity to be measured and the measuring apparatus
(observer).~\footnote{This is the main idea behind the ``relational'' approach to (quantum) physics
(see~\cite{Rovelli:2021thu} and references therein).}
The prototype is given by the energy of a particle, which is obtained by the contraction of the particle's
4-velocity $u^\mu$ with the observer's 4-velocity $U^\mu$, that is $E=-g_{\mu\nu}\,u^\mu\,U^\nu$,
where $g_{\mu\nu}$ is the background metric tensor.
Similarly, one might notice that fields in configuration space have no direct physical meaning.
This makes fields like spacetime coordinates but also as tensor quantities in spacetime: proper observables,
like scattering cross sections, are built out of fields by (more or less) involved ``scalarizations'' involving the measuring apparatus.
\section{Background independence and field redefinitions}
\label{back-red}
\setcounter{equation}{0}
At the level of the naive effective action $\Gamma^{(0)}[\phi] \equiv \Gamma_\text{N}[\phi]$, no absolute structure exists,
thus $\Sigma = \varnothing$ and the covariance \eqref{eq:cov} collapses into invariance~\eqref{eq:inv}.
However, like we saw in Section~\ref{sec:EA}, $\Gamma^{(0)}[\phi]$ depends on the field parameterization and, therefore,
does not satisfy invariance~\eqref{eq:inv}.

This can be amended by the introduction of a configuration-space metric $G_{ij}$, which is used to define an invariant
path-integral measure and ultimately results in the Vilkovisky-DeWitt effective action
$\Gamma^{(1)}[\phi, \Sigma] \equiv \Gamma_\text{VD}[\phi, G]$.
Although $\Gamma^{(1)}[\phi, \Sigma]$ satisfies invariance \eqref{eq:inv}, we now have an absolute structure
$\Sigma = \{G_{ij}\}$ that renders the theory background dependent. 

Let us make the theory background independent by promoting $G_{ij}$ to a dynamical structure.
On assuming ultralocality,
\begin{equation}
	G_{ij}
	= G_{IJ} \, \delta(x,x')
	\ ,
\end{equation}
one could envisage doing so by introducing the action
\begin{equation}
	S = \int\mathrm{d}^n x \sqrt{-g} \, \mathcal{L}(G_{IJ},R_{IJMN})
	\ ,
\end{equation}
where $R_{IJMN}$ is the Riemann tensor of the finite-dimensional space $\mathcal N$ and $\mathcal{L}(G_{IJ},R_{IJMN})$
is some Lagrangian invariant under spacetime and configuration-space diffeomorphisms.
The precise form of the action for $G_{IJ}$ is not essential to our argument as the only crucial point to the following
is the existence of dynamics (regardless of the particular kind of evolution).
One could, for example, assume that such a Lagrangian takes the Einstein-Hilbert form
\begin{equation}
	\mathcal{L}(G_{IJ},R_{IJMN})
	= G^{IJ} R_{IJ}
	\ ,
\end{equation}
where $R_{IJ} = G^{MN} R_{MINJ}$ is the configuration-space Ricci tensor.

With a dynamical configuration space, there is no other absolute structure present,
thus the theory recovers its background independence $\Sigma = \varnothing$.
However, this came at the cost of introducing the additional degree of freedom $G_{ij}$
in the theory, which must be quantised along with $\varphi^i$.
In the absence of external sources, the quantum theory then reads
\begin{equation}
\exp \left( \frac{i}{\hbar}\, \Gamma^{(2)}[\phi] \right)
=
\int\mathcal{D}\phi\mathcal{D}G_{ij}\,
\sqrt{\Det G_{ij}}
\exp\left(
\frac{i}{\hbar}\,
S[\phi,G]
\right)
\ ,
\label{eq:qadef}
\end{equation}
The functional measure
\begin{equation}
	\mathcal{D}\phi\,\mathcal{D}G_{ij} \,\sqrt{\Det G_{ij}}
	=
	\prod_{x,\, I<J} \d\phi \, \d G_{IJ} \sqrt{\det G_{IJ}}
\end{equation}
does not transform covariantly. To see this, it is sufficient to perform an infinitesimal field redefinition
\begin{equation}
	G_{ij} \to G'_{ij} = G_{ij} + \xi_{ij}(G)
	\ ,
\end{equation}
under which the functional measure transforms as
\begin{equation}
	\mathcal{D}G'_{ij} \,\sqrt{\Det G'_{ij}}
	=
	\mathcal{D}G_{ij} \,\sqrt{\Det G_{ij}}
	\left(1 + G^{mn}\,\xi_{mn} + \frac{\delta \xi_{mn}}{\delta G_{mn}}\right)
	\ .
\end{equation}
Promoting $G_{ij}$ to a dynamical field thus brings back the violation of invariance under field redefinitions.
To correct this, we must find a different measure that transforms covariantly.
This unavoidably requires the introduction of some absolute structure $\mathcal S^{(2)}$
that transforms in such a way that~\footnote{Should the geometrical construction be maintained,
$\mathcal S^{(2)} = \sqrt{\Det G^{(2)}_{i_2j_2} }$ would correspond to the determinant of a second-order
functional metric $G^{(2)}_{i_2j_2}$ defined on a second-order configuration space $\mathcal{C}^{(2)}(M)$,
where now the indices $i_2,j_2$ include the new field $G_{ij}$.}
\begin{equation}
	\mathcal{D}\phi\,\mathcal{D}G'_{ij}\, \sqrt{\Det G'_{ij}} \, \mathcal S^{(2)'}
	=
	\mathcal{D}\phi\,\mathcal{D}G_{ij} \,\sqrt{\Det G_{ij}} \, \mathcal S^{(2)}
	\ ,
\end{equation}
which recovers background dependence since $\Sigma = \{\mathcal S^{(2)}\}$.
We thus see that promoting $\mathcal S^{(n)}$ to a dynamical field requires, upon quantization,
that an absolute structure $\Sigma = \{\mathcal S^{(n+1)}\}$ be introduced, hence the effective
action $\Gamma^{(n)}[\phi, \Sigma]$ can never satisfy invariance~\eqref{eq:inv} and background
independence $\Sigma = \varnothing$ at the same time.

In summary, although we have been able to recover background independence by making $G_{ij}$
dynamical, we lost invariance in the full quantum theory where $G_{ij}$ belongs to the spectrum.
To regain invariance we would need to include another layer of abstraction (a sort of 
{\em \"uber}-field space) with yet another
higher-order absolute structure, which would then face the problems with background dependence
and so on and so forth.
This process continues \textit{ad infinitum}, proving the existence of a tension between background
independence and covariance in quantum gravity.
\section{Conclusions}
\label{sec:conc}
\setcounter{equation}{0}
In this paper, we have considered extended notions of diffeomorphism invariance and
background independence related to the (non) existence of some fixed metric structure
at the quantum level.
In this context, diffeomorphism invariance also comprises invariance of observables under field redefinitions,
which are required to enforce gauge independence in the correlation functions.
Likewise, a metric structure in field space is naturally introduced by any given dynamical
theory.
\par
We have then shown that diffeomorphism invariance and background independence, in the
extended sense of the absence of any fixed metric structure at any level, cannot coexist.
While the former is unavoidable for any sensible physical theory, the latter does not seem to be rooted
in any physical property other than an aesthetic appeal (to some).
Therefore, one possible and natural implication of our finding is that quantum gravity is actually
background dependent (at some level in the hierarchy of mathematical structures
described at the end of the last section).
We stress that our definition of background independence concerns only (local)
geometrical objects, such as metric structures in the \textit{\"uber}-field spaces.
As we already stated, points, dimension, topological and differential structures are not included in this definition.
On the other hand, should background independence be revealed as an intrinsic aspect of Nature,
models of quantum gravity based on quantum field theory, such as asymptotic safety and causal
dynamical triangulation, would become effective rather than fundamental theories.
\section*{Acknowledgments}
R.C.~and A.K. are partially supported by the INFN grant FLAG.
The work of R.C.~has also been carried out in the framework of activities of the National Group of Mathematical Physics
(GNFM, INdAM).

\begin{thebibliography}{10}
%
\bibitem{Norton_1993}
J.~D.~Norton,
Rep. Prog. Phys. \textbf{56} 791 (1993) 791.
%
\bibitem{Pooley:2015nfa}
O.~Pooley,
Einstein Stud. \textbf{13} (2017) 105
[arXiv:1506.03512 [physics.hist-ph]].
%
\bibitem{Anderson}
J.~L.~Anderson,
``Principles of Relativity Physics'',
Academic Press,
1967.
%
\bibitem{Giulini:2006yg}
D.~Giulini,
Lect. Notes Phys. \textbf{721} (2007) 105
[arXiv:gr-qc/0603087 [gr-qc]].
%
\bibitem{Colosi:2004sa}
D.~Colosi, L.~Doplicher, W.~Fairbairn, L.~Modesto, K.~Noui and C.~Rovelli,
Class. Quant. Grav. \textbf{22} (2005) 2971
[arXiv:gr-qc/0408079 [gr-qc]].

\bibitem{Smolin:2005mq}
L.~Smolin,
``The Case for background independence,''
[arXiv:hep-th/0507235 [hep-th]].

\bibitem{Polchinski:1998rq}
J.~Polchinski,
``String theory. Vol. 1: An introduction to the bosonic string,''
Cambridge University Press, 2007.
\bibitem{Polchinski:1998rr}
J.~Polchinski,
``String theory. Vol. 2: Superstring theory and beyond,''
Cambridge University Press, 2007.
\bibitem{Witten:1993ed}
E.~Witten,
``Quantum background independence in string theory,''
[arXiv:hep-th/9306122 [hep-th]].

\bibitem{Rozali:2008ex}
M.~Rozali,
Adv. Sci. Lett. \textbf{2} (2009) 244
[arXiv:0809.3962 [gr-qc]].
%

\bibitem{DeWitt:1967yk}
B.~S.~DeWitt,
Phys. Rev. \textbf{160} (1967) 1113.
%
\bibitem{Bergmann:1966zza}
P.~G.~Bergmann,
Phys. Rev. \textbf{144} (1966) 1078.
%
\bibitem{Dirac:1958sq}
P.~A.~M.~Dirac,
Proc. Roy. Soc. Lond. A \textbf{246} (1958) 326;
%
P.~A.~M.~Dirac,
``Lectures on quantum mechanics,''
Dover, 1964.
%
\bibitem{kiefer}
C.~Kiefer,
``Quantum Gravity,''
Oxford University Press, 2007.

\bibitem{Rovelli:2004tv}
C.~Rovelli,
``Quantum gravity,''
Cambridge University Press, 2004.
%
\bibitem{Ashtekar:2004eh}
A.~Ashtekar and J.~Lewandowski,
Class. Quant. Grav. \textbf{21} (2004) R53
[arXiv:gr-qc/0404018 [gr-qc]].
%
\bibitem{Abbott:1981ke}
L.~F.~Abbott,
Acta Phys. Polon. B \textbf{13} (1982) 33.
%
\bibitem{reuter}
M.~Reuter,
Phys. Rev. D \textbf{57} (1998) 971
[arXiv:hep-th/9605030 [hep-th]];
M.~Becker and M.~Reuter,
Phys. Rev. D \textbf{102} (2020) 125001
[arXiv:2008.09430 [gr-qc]];
M.~Becker and M.~Reuter,
Phys. Rev. D \textbf{104} (2021) 125008
[arXiv:2109.09496 [hep-th]].
%
\bibitem{Vilkovisky:1984st}
G.~A.~Vilkovisky,
Nucl. Phys. B \textbf{234} (1984) 125.
%
\bibitem{DeWitt:1988dq}
B.~S.~DeWitt,
``The effective action'',
in Batalin, I.A. (Ed.) et al.: Quantum field theory and quantum statistics, Vol. 1, 191-222,
CRC Press (1987).

\bibitem{Wagoner}
R.W. Wagoner, Phys. Rev. D 1 (1970) 3209-3216

\bibitem{Kam-Stein}
A.Yu. Kamenshchik and C.F. Steinwachs, Phys. Rev. D 91 (2015) 8, 084033 [arXiv: 1408.5769 [gr-qc]]

\bibitem{Falls:2018olk}
K.~Falls and M.~Herrero-Valea,
Eur. Phys. J. C \textbf{79} (2019) 595
[arXiv:1812.08187 [hep-th]].
%

\bibitem{Gianni}
A.Yu. Kamenshchik, E.O. Pozdeeva, A. Tronconi, G. Venturi and S.Yu. Vernov,
Class. Quant. Grav. 31 (2014) 105003 [arXiv: 1312.3540 [hep-th]]

%
\bibitem{Rebhan:1987cd}
A.~Rebhan,
Nucl. Phys. B \textbf{298} (1988) 726.

\bibitem{BK}
P.~G.~Bergmann and A.~Komar,
Int. J. Theor. Phys. \textbf{5} (1972) 15.

%
\bibitem{Casadio:2021rwj}
R.~Casadio, A.~Kamenshchik and I.~Kuntz,
Nucl. Phys. B \textbf{971} (2021) 115496
[arXiv:2102.10688 [hep-th]].

\bibitem{Ehlers1995}
J.~Ehlers,
``Machian Ideas and General Relativity'',
in Barbour, J.B. (Ed.) et al.: Mach's principle - from Newton's bucket to quantum gravity,
458-473, Birkh\"auser (1995).
%
\bibitem{Einstein1922}
A.~Einstein,
``The Meaning of Relativity'',
Princeton University Press (1922).

%
\bibitem{Landsman:2022caj}
K.~Landsman,
``Reopening the Hole Argument,''
[arXiv:2206.04943 [gr-qc]].
%
\bibitem{Rovelli:2021thu}
C.~Rovelli,
``The Relational Interpretation of Quantum Physics,''
[arXiv:2109.09170 [quant-ph]].
%
%
\end{thebibliography}
\end{document}